\documentclass[conference]{IEEEtran}
\usepackage{amsmath}
\usepackage{amssymb}
\usepackage{lipsum}
\usepackage[font=footnotesize, labelfont=bf]{caption}
\usepackage{listings}
\usepackage{gensymb}
\usepackage{fancyhdr}
\usepackage{graphicx}
\usepackage[nocompress]{cite}

\begin{document}
\title{Potential benefits of a block-space GPU approach for discrete tetrahedral domains}
\author{
    \IEEEauthorblockN{Crist\'obal A. Navarro}
    \IEEEauthorblockA{Instituto de Inform\'atica\\
                        Universidad Austral de Chile\\
                        Valdivia, Chile}
    \and
    \IEEEauthorblockN{Benjam\'in Bustos}
    \IEEEauthorblockA{Departamento de Ciencias de la Computaci\'on)\\
                        Universidad de Chile\\
                        Santiago, Chile}
    \and
    \IEEEauthorblockN{Nancy Hitschfeld}
    \IEEEauthorblockA{Departamento de Ciencias de la Computaci\'on\\
                        Universidad de Chile\\
                        Santiago, Chile}
}

\maketitle
\begin{abstract}
The study of data-parallel domain re-organization and thread-mapping techniques
are relevant topics as they can increase the efficiency of GPU computations when
working on spatial discrete domains with non-box-shaped geometry.  In this work we
study the potential benefits of applying a succint data re-organization of a
tetrahedral data-parallel domain of size $\mathcal{O}(n^3)$ combined with an
efficient block-space GPU map of the form $g(\lambda):\mathbb{N} \rightarrow
\mathbb{N}^3$. Results from the analysis suggest that in theory the combination
of these two optimizations produce significant performance improvement as
block-based data re-organization allows a coalesced one-to-one correspondence at
local thread-space while  $g(\lambda)$ produces an efficient block-space spatial
correspondence between groups of data and groups of threads, reducing the number
of unnecessary threads from $O(n^3)$ to $O(n^2\rho^3)$ where $\rho$ is the
linear block-size and typically $\rho^3 \ll n$. From the analysis, we obtained 
that a block based succint data re-organization can provide up to $2\times$ improved
performance over a linear data organization while the map can be up to 
$6\times$ more efficient than a bounding box approach. The results from this
work can serve as a useful guide for a more efficient GPU computation on
tetrahedral domains found in spin lattice, finite element and special n-body
problems, among others.
\end{abstract}
\IEEEpeerreviewmaketitle

\section{Introduction}
GPU computing has proven to be both a practical tool and a well established
research area for computer science \cite{4490127,
Nickolls:2010:GCE:1803935.1804055, navhitmat2014}, mostly because there are
several parallel computing issues that get magnified when handling thousands of
processors in parallel and become critical for achieving high efficiency. One of
these problems is related to how memory can be accessed in parallel as it is
no longer possible to assume an exclusive data path for each processor, but
instead one must assume that one memory access can provide several consecutive
words of data to a group of threads.  A more recent problematic is related to
the challenge of \textit{given a data-parallel problem, find a thread map that
uses a space of computation that is close to the optimal one}, where optimal is
defined as the space of computation that achieves maximum parallelism with all
threads doing useful work. This problem typically arises when handling fine
grained data-parallel problems defined on a \textit{complex spatial domain},
\textit{i.e.,} one that is different from the box-shape for the corresponding
spatial dimensions of the problem.

Since the introduction of general purpose GPU programming tools such as CUDA
\cite{nvidia_cuda_guide} and OpenCL \cite{opencl08}, the problems recently
described have gained mayor importance as they indeed appear in the programming
model and have an impact on the performance of the GPU. Although these
problems exist in all \textit{complex spatial domains}, in this work we are
particularly interested in studying the potential benefits on 3D discrete
triangular domains as they are found in important applications such as special
\textit{all-with-all} problems, computational physics on spin lattices, special
n-body problems and cellular automata under special boundary conditions.

In the CUDA GPU programming model there is a hierarchy of three
constructs\footnote{OpenCL chooses different names for these constructs; (1)
work-element, (2) work-group and (3) work-space, respectively.} that are defined
for the execution of a highly parallel algorithm; (1) thread, (2) block and (3)
grid.  Threads are the smallest elements and they are in charge of executing the
instructions of a GPU kernel.  A block is an intermediate structure that
contains a set of threads organized in an Euclidean space.  Blocks provide fast
shared memory access as well as synchronization for all of its threads. The grid
is the largest construct of the GPU and it is in charge of keeping all blocks
spatially organized. These three constructs play an important role when mapping
the execution resources to the problem domain as well as for the memory accesses.  

The problems already mentioned can be described in more detail for the case of a
discrete 3D triangular domain, where elements have a spatial organization:
\begin{enumerate}
    \item A typical linear memory organization of data elements in a 3D discrete
        pyramid in a major depth-row order, \textit{i.e.,} $z \rightarrow y
        \rightarrow x$, leads to a non-linear pattern of linear distances
        between nearest neighbors. This aspect produces a negative impact on
        performance as coalesced memory accesses become less frequent.
    
    \item There is a stage in the GPU computing pipeline where the \textit{space
        of computation} is mapped to the problem domain. This map can be
        defined as a function $f(x): \mathbb{N}^k \rightarrow \mathbb{N}^q$ that
        transforms each $k$-dimensional point $x=(x_1, x_2, ..., x_k)$ of the
        grid to a unique $q$-dimensional point of the problem domain. When the
        problem domain is simple in shape, \textit{e.g.,} a box, the canonical 
        GPU map $f(x) = x$ becomes the optimal and simplest one.
        For the case of 3D discrete triangular domains, up to $O(n^3)$ threads
        may become unnecessary with this approach, therefore it is of interest
        to find a more efficient approach that can map threads according to a 3D
        triangular distribution.
\end{enumerate}

In this work we analyze the potential performance benefits of applying a 
fast re-organization of the problem domain combined with a block-space map that 
preserves thread locality. Prior works on this subject are considered and cited
in the next Section. 

\section{Related Work}
\label{sec_relatedwork}
Wu \textit{et. al.} proved that the problem of data re-organization for parallel 
computing is \textit{NP-Complete} in its general form 
\cite{Wu:2013:CAA:2517327.2442523}, nonetheless the authors describe efficient 
approaches based on data replication, padding and sharing and indicate which
ones fit better for certain problem categories. Chen \textit{et. al.}
propose a general optimization technique for data-parallel problems with
indirect memory accesses \cite{Chen:2016:ERS:2854038.2854046}, by viewing 
the problem as a sparse matrix computation. Yavors'kii and Weigel identified
that tiled computations, such as the ones found in spin systems, are greatly 
improved by re-organizing the memory in blocks \cite{Yavors’kii2012}. 

Regarding thread mapping techniques, Jung \textit{et. al.} \cite{Jung2008}
proposed packed data structures for representing triangular and
symmetric matrices with applications to LU and Cholesky decomposition
\cite{springerlink_gustavson}. The strategy is based on building a
\textit{rectangular box strategy} for accessing and storing a triangular
matrix (upper or lower). 
Ries \textit{et. al.} contributed with a parallel GPU method for the
triangular matrix inversion \cite{Ries:2009:TMI:1654059.1654069}.  The authors
identify that the space of computation indeed can be improved by using a
\textit{recursive partition} of the grid, based on a \textit{divide and
conquer} strategy.
Navarro and Hitschfeld studied the benefits of block-space thread mapping for 2D
triangular domains \cite{DBLP:conf/hpcc/NavarroH14} and found approximately
$20\%$ of improvement for 2D triangular problems such as collision detection and
the Euclidean distance Matrix.

The study of discrete 3D triangular structures is an interesting category of
problem with a variety of applications, from spin lattice simulation, special
triplet collision and 3D Euclidean distance matrices. For such applications, 
a data re-organization technique and block-space thread mapping can provide
a substantial performance improvement.

\section{Analysis of the Block-Space Approach}
We consider the 3D pyramid case that is defined by $n$ triangular structures
stacked and aligned at their middle corner, where the $n$-th triangle contains $T_n^{2D} =
n(n+1)/2$ elements. The total number of elements for the full structure is
\begin{equation}
    T_n = \sum_{i=1}^{n}{\frac{i(i+1)}{2}}
\end{equation}
The sequence corresponds to the tetrahedral numbers, which are
defined by 
\begin{equation}
    T_n = {{n+2}\choose{3}} = \frac{n(n+1)(n+2)}{6}
\end{equation}

The following analysis combines a simple yet effective succinct data
re-organization scheme with a map of the form $g(\lambda):\mathbb{N} \rightarrow
\mathbb{N}^3$, both working in block-space.

\subsection{Succinct block re-organization scheme}
Data layout for GPU computing typically follows a linear memory approach as it
is often a copy of the data layout found in the host side. For spatial nearest
neighbor GPU computations on a 3D triangular domain, a succinct block-based
re-organization of data becomes an attractive optimization to be considered as
it provides clean coalesced memory access for all threads. 

Let $M$ be the memory space, $A_k$ an alignment of length $k$ measured in bytes
and $\omega$ the size of a warp of threads (typically $\omega = 32$). An
analysis on the access patterns of a warp is sufficient to model, in great part,
the efficiency of GPU memory accesses on the 3D triangular structure. A warp may
access memory with an offset of $\Delta$ from the alignment and a stride of $s$
bytes. When a warp's memory access is aligned to $A_k$ and the stride is $s=0$,
the number of memory accesses required corresponds to $\omega b/k$ bytes where
$b$ is the number of bytes accessed by each thread.  Typically, the number of
consecutive bytes read by a warp matches the $k$-bytes transaction, making the
operation cost just one memory access for the whole warp. 

A linear organization of the pyramid in $M$ produces a non-linear pattern for
the distances among data elements.  That is, for any data element $d_{x,y,z}$
with linear memory coordinate $\lambda_d$, the linear memory distances to its
top and bottom nearest neighbors, $\delta(d_{x,y,z}, d_{x,y+1,z})$ and
$\delta(d_{x,y,z}, d_{x,y-1,z})$, vary for each different variation of the
coordinates $y,z$ in the pyramid, producing at least one extra memory access for
each misaligned warp.  In order to count the number of misaligned warps in a
pyramid, we consider the case of a single triangular layer of size $T_n^{2D} =
n(n+1)/2$ elements and then extend the result to the 3D pyramid. 

Given
an alignment $A_k$, the number of rows aligned to $k$-bytes is
\begin{equation}
    R_{k,n} = \Bigg\lfloor \frac{n}{k + k((k+1)\mod 2)}  \Bigg\rfloor 
\end{equation}
Since alignments use even values of $k$, $R_{k,n}$ becomes 
\begin{equation}
    R_{k,n} = \Bigg \lfloor \frac{n}{2k} \Bigg \rfloor
\end{equation}
where the total number of warps aligned in one 2D triangular layer of side $n$
is defined and upper bounded as
\begin{equation}
    W_{k,n} = \sum_{i=1}^{R_k} 2i = R_k(R_k+1) \le n^2/4k^2 + n/2k.
\end{equation}
The number of aligned warps $W_k$ decreases considerably as $k$ increases.
Moreover, the fraction of aligned warps can be no greater than
\begin{equation}
    F_{A_k,n} = \frac{W_k}{T_n^{2D}/k} = \frac{W_k}{\lceil n(n+1)/2k \rceil} < \frac{1}{2k} + \frac{1}{n}
\end{equation}
For an alignment of $k=128$ bytes, which is a common case where each one of the
32 threads of a warp accesses a single float of 4 bytes with $s=0$, the total
percentage of aligned warps would be no greater than $F_{A_{128}} \le 0.39\% +
1/n$. Considering that the pyramid corresponds to stacked layers of size
$1,2,\dots, n$ where all present the same behavior, if not more complex, we
expect the total cost for accessing all data once to be at least
\begin{equation}
    C(\alpha, k, n) = \frac{T_n}{k}  \Big(F_{A_k, i} + \alpha (1 - F_{A_k,i})\Big)
\end{equation}
where $\alpha$ is a cost defined for an unaligned memory access. In the best possible
scenario, the cost of an uncoalesced access would incur in at least one extra operation,
\textit{i.e.,} $\alpha = 2$, which would lead to a cost of
\begin{equation}
    C(\alpha, k, n) = \frac{T_n}{k} \Big(2 - F_{A_k,i}\Big)
\end{equation}

A succinct blocked re-organization of the pyramid can produce a different cost
function with full alignment of warps with data. At a coarse level, the
structure can be represented by blocks of data linearly organized in $M$.  At a
fine-grained level, each block has constant size $\Theta(\rho^3)$ with a local linear
organization and with $\rho = k$ to match the alignment.  For the elements of
the diagonal region, blocks are padded to preserve memory alignment for the rest
of the structure.  This design leads to a succinct blocked structure of asymptotic
size $O(n^3) + o(n^3)$ with $F_{A_k} = 1$, which leads to a cost of
\begin{equation}
    C'(\alpha, k, n) = \frac{T_n + n^2\rho^3}{k}
\end{equation}
where typically $\rho^3 \ll n$. For large $n$ and $\alpha = 2$, the potential
improvement factor for data re-organization becomes
\begin{equation}
    \frac{C(2, k, n)}{C'(2, k, n)} = \frac{2T_n -
        T_nF_{A_k}}{T_n + n^2\rho^3} \approx 2 - F_{A_k} \le 2
\end{equation}
Based on the possibilities of improvement and considering that in practice $F_{A_k}$ is 
a low value for the pyramid case, it is highly convenient to re-organize the 
data of a pyramid, from a linear scheme to a succinct block-based one.

\subsection{Block-based thread map}
The use of a box-shaped grid to map threads on a 3D domain is a
standard approach used in GPU computing and the natural one provided 
by the computing model since it is effective and efficient for many
data-parallel problems. However the strategy presents a strong inefficiency when
dealing with non box-shaped domains such as the pyramid as the number of
unnecessary threads is in the order of $O(n^3)$. A more efficient approach can
be used by considering how indices are organized in a pyramid.  

It is possible to use a map of the form $g(\lambda):\mathbb{N}
\rightarrow \mathbb{N}^3$ that uses reduced set of blocks that are mapped directly to the
pyramidal structure without loss of parallelism. The approach takes advantage of
the fact that when using a linear enumeration of blocks on the pyramid, the
linear index $\lambda$ of the first element of a 2D triangular layer corresponds
to a tetrahedral number $T_n$. Based on this fact, the rest of the data elements
that reside in the same layer must follow the property:
\begin{equation}
    \sum_{k+1}^{z} {k(k+1)/2} < \lambda = \sum_{k=1}^v {k(k+1)/2} < \sum_{k=1}^{z+1} {k(k+1)/2}
\end{equation}
Considering the expression for the tetrahedral numbers, we have that
\begin{equation}
    \lambda = \sum_{k=1}^v {k(k+1)/2} = \frac{v(v+1)(v+2)}{6}
\end{equation}
therefore, given the linear location $\lambda$ of a block, one can obtain its
$z$ coordinate in the pyramid by solving the equation:
\begin{equation}
    v^3 + 3v^2 + 2v -6\lambda = 0
\end{equation}
and extracting the integer part of the root
\small
\begin{equation}
    v   = \frac{\sqrt[3]{\sqrt{729\lambda^2 - 3} + 27\lambda}}{3^{2/3}} + \frac{1}{\sqrt[3]{3} \sqrt[3]{ \sqrt{729\lambda^2 - 3} + 27\lambda}} - 1
\end{equation}
\normalsize
Once the value $z = \lfloor v \rfloor$ is computed, one can obtain the
two-dimensional $\lambda'$ linear coordinate
\begin{equation}
    \lambda' = \lambda - T_z 
\end{equation}
where $T_z = z(z+1)(z+2)/6$ is the tetrahedral number for the recently computed
$z$ value. With $\lambda'$ already computed, one can obtain the $x$ and $y$
values of the block by using the triangular map proposed by Navarro and
Hitschfeld \cite{DBLP:conf/hpcc/NavarroH14} for 2D triangular domains. Combining
all three computations, map $g(\lambda)$ becomes
\begin{equation}
    g(\lambda) \mapsto (x,y,z) = \Big(\lambda' - T_y^{2D}, \Big\lfloor\sqrt{\frac{1}{4} +
2\lambda'} - \frac{1}{2}\Big\rfloor, \lfloor v \rfloor\Big)
\end{equation}
where $T_y^{2D}$ is the triangular number for $y$.

The block linear size is defined as $\rho = k$ to match the data
re-organization scheme. For practical purposes, the blocks can be organized 
on a cubic grid of $\sqrt[3]{T_n}$ in order to balance the number of elements on
each dimension, producing $n^2 \rho^3$ unnecessary threads which correspond to
the succint data. For the thread mapping stage, where the cost of an unnecessary
thread is comparable as to a worker thread, one can write the potential
improvement factor of the pyramidal with respect to the box strategy as 
\begin{equation}
I = \frac{\beta n^3/\rho^3}{\tau T_n/\rho^3} = \frac{6\beta n^3}{\tau (n^3 + 3n^2 + 2n)}
\end{equation}
where $\beta$ is the cost of computing the block coordinate using the box
approach, while $\tau$ is the cost mapping blocks in the pyramidal map. 
In the infinite limit of $n$, the potential improvement becomes
\begin{equation}
    I_{n\to\infty} \sim \frac{6\beta}{\tau}
\end{equation}
and tells that in theory the pyramidal map could be up to $6\times$ faster
than the box approach. However, the improvement observed in experimentation will 
depend on how efficient is $\tau$ compared to $\beta$, \textit{i.e.,} how
efficient are the cubic and square root computations performed.

\section{Conclusions}
The optimization techniques analyzed in this work can offer significant
potential improvement that are worth considering for future GPU computations on
special spin systems, cellular automata, n-body problems and any other problem
for which is useful to consider the pyramidal domain. In theory, it is possible
to extract up to $2\times$ more performance from a simple succinct data
re-organization and be up to $6\times$ more efficient by using an specialized
pyramidal map.  However, the effective improvement observed in practice will
strongly depend on how much overhead is necessarily introduced when
re-organizing data as well as how expensive will the cubic and square root
computations become in practice.  As a future work, it will be interesting to
consider technical optimizations for the GPU architecture in order to obtain an
experimental performance that represents the theoretical results obtained.
\bibliographystyle{plain}
\bibliography{bare_conf}

\end{document}